\documentclass[aps,pra,reprint,superscriptaddress]{revtex4-1}
\usepackage{graphicx}
\usepackage{amsmath}
\draft % marks overfull lines with a black rule on the right
 \usepackage{bm}
 \usepackage{mathrsfs}
\usepackage{color}
\usepackage{float}
 \usepackage{bm}
 \usepackage{mathrsfs}
 \usepackage{amsmath}

\begin{document}

%%%%%%%%%%%%%%%%%% title page information %%%%%%%%%%%%%%%%%%
\title{Fano lines in the reflection spectrum of directly coupled systems of waveguides and cavities: measurements, modeling and manipulation of the Fano asymmetry}

\author{Jin Lian$^{1,2,*}$, Sergei Sokolov$^{1,2}$, Emre Y\"uce $^{2,3}$, Sylvain Combri\'e$^{4}$,  Alfredo De Rossi$^{4}$, and Allard P. Mosk }

\affiliation{
Nanophotonics, Debye Institute for Nanomaterials Science, Center for Extreme Matter and Emergent Phenomena, Utrecht University, P.O. Box 80.000, 3508 TA  Utrecht, The Netherlands \\
$^2$Complex Photonic Systems (COPS), MESA+ Institute for
Nanotechnology, University of Twente, P.O. Box 217, 7500AE  Enschede, The Netherlands \\
$^3$Light $\&$ Matter Control Group, Department of Physics,Middle East Technical University, 06800 Ankara, Turkey\\
$^4$Thales Research and Technology, Route D\'epartementale 128, 91767 Palaiseau, France\\
$^*$Corresponding author: IamJinLian@gmail.com
}


\makeatletter
\let\setrefcountdefault\relax
\makeatother
%\date{SS141010 \today}
\date{\today}

\begin{abstract}
We measure and analyze reflection spectra of directly coupled systems of waveguides and cavities. The observed Fano lines offer insight in the reflection and coupling processes. Very different from side-coupled systems, the observed Fano line shape is not caused by the termini of the waveguide, but the coupling process between the measurement device fiber and the waveguide. Our experimental results and analytical model show that the Fano parameter that describes the Fano line shape is very sensitive to the coupling condition. A movement of the fiber well below the Rayleigh range can lead to a drastic change of the Fano line shape. 

\end{abstract}

%\ocis{(130.5296) Photonic crystal waveguides, (260.5740) Resonance.} % REPLACE WITH CORRECT OCIS CODES FOR YOUR ARTICLE, MINIMUM OF TWO; Avoid using the OCIS codes for “General” or “General science” whenever possible.

%%%%%%%%%%%%%%%%%%%%%%% References %%%%%%%%%%%%%%%%%%%%%%%%%
\maketitle %\maketitle must follow title, authors, abstract and \pacs

%\section{}
 %\label{}
%\subsection{}
%\subsubsection{}
\noindent 

%%%%%%%%%%%%%%%%%%%%%%%%%%  body  %%%%%%%%%%%%%%%%%%%%%%%%%%
\section{Introduction}
Photonic crystal (PhC) \cite{mitbook} cavities are of tremendous interest for device applications due to their beneficial properties such as small mode volume and high quality factor ($Q$) \cite{Kuramochi2006,Tanabe2007,Song2005}. Multiple-cavity systems are great platforms to study fundamental physics and build all-optical devices \cite{Yariv1999,Hafezi2013,Notomi2008,Xia2007,Hartmann2008,Yanik2004}. Thus, the characterization of multiple-cavity systems composed of photonic crystal cavities is of great importance.  Reflection measurement is a typical way of characterizing resonant systems. From a reflection spectrum, the resonance width and frequency can be obtained. Depending on the structure of a system, resonances shown in a spectrum can be Lorentzian shape or Fano line shape \cite{Fano1961}.

The Fano line shape shows up when a narrow resonance interferes with a continuum. Depending how the continuum interacts with the resonance, a Fano line can show up various profiles \cite{Fano1961,Miroshnichenko2010}. In many cases, the sharp asymmetric Fano line shape is preferred to the Lorenztian shape. For example, for optical switching Fano resonances reduce the switching thresholds and give much larger switching contrast \cite{Yang2007, Nozaki2013}. Fano lines appear very often and are widely studied in side-coupled single and multi-cavity and waveguide systems.  For side-coupled single cavity and waveguide systems,  Fano lines can be created by adding extra scattering or reflecting elements in waveguides since transmission is open within the bandwidth of the waveguides \cite{Fan2002, Heuck2013, Yu2014}.  For side-coupled multi-cavity and waveguide system, Fano line shapes show up naturally due to the direct and indirect cavity-cavity couplings even without extra scattering and reflecting elements \cite{lin2005, zhou2014} .

The situation is different in directly coupled single and multi-cavity and waveguide systems (Fig.1). In such systems transmission is only open at the cavity resonance, and indirect cavity-cavity couplings are absent.  In case the light frequency is far off the resonance of the PhC cavity, light will be completely reflected, since the system is closed. Therefore, reflection of the facet of the PhC waveguide does not drastically change the line shape.  Only in case the frequency of the incident light is close to the resonance of the PhC cavity, the reflection of the waveguide plays a role. In Fig. 2, we show the calculated spectra of direct-coupled multi-cavity and waveguide systems in two different cases, one with the consideration of the reflection of the waveguide, the other without taking into account of the reflection of the waveguide. In both cases, we see symmetric Lorentzian shapes.  In contrast, we observe sharp asymmetric line shapes on top of Fabry-P\'erot fringes in our experiments. Apparently, the observed strong asymmetric Fano line shape in a directly coupled systems can not be solely attributed to the termini of the waveguide.  

\begin{figure}[H]
\centerline{\includegraphics[width=0.99\columnwidth]{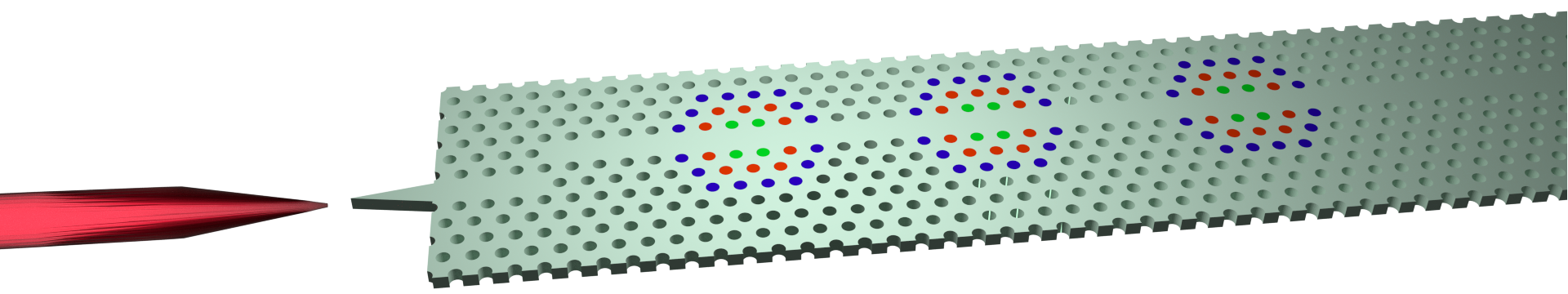}}
\caption{A schematic representation of the sample and experimental setup. The green membrane structure represents the sample. In the barrier waveguide (middle line defect), three cavities have been created by shifting the holes around the waveguided in a tapered way. The shift of the green holes are $S1=0.0124a$, the shifts of the red and blue holes are 2/3 $S1$ and 1/3 $S1$, and $a$ is the lattice constant. The red cone next to the sample represents a polarization maintaining lensed fiber (PMF) used to couple light into the membrane. }
\end{figure}

%As discussed before, very different from side coupled system,  the reflection of the facet of the waveguide does not lead to asymmetric Fano line shapes. 
%Therefore, the observed Fano line shapes in our experiment are not trivial, and its physical cause of them needs to be uncovered. 

The possibility to tune the asymmetry of the Fano line shape also arouses much interest. This tuning has been achieved  by tuning the cavity resonances \cite{Fan2002, Xiao2009,Verhagen09,Frimmer2012}. For a given cavity, by deliberate design of the coupling to the continuum, different Fano line shapes can be obtained by fabricating different samples \cite{Chang2013,Yu2014, Zhang2016}. However, after the structure is fabricated, it is very difficult to change the Fano line shape without changing the frequency of the resonance because this requires to change the properties of the continuum or the broad resonances.

In this work, we experimentally and theoretically investigate Fano resonances in a  multiple-cavity system directly coupled to a waveguide in a PhC membrane structure.  We measure the Fano lines and manipulate their shape by tuning the resonances.  We create an analytical model which uncovers the origin of the Fano line shape, and it accurately reproduces our experimental results. With the help of our model, we propose and experimentally demonstrate a way of directly manipulating the Fano lineshape without tuning the cavity resonances.  

\begin{figure}[H]
\centerline{\includegraphics[width=0.9\columnwidth]{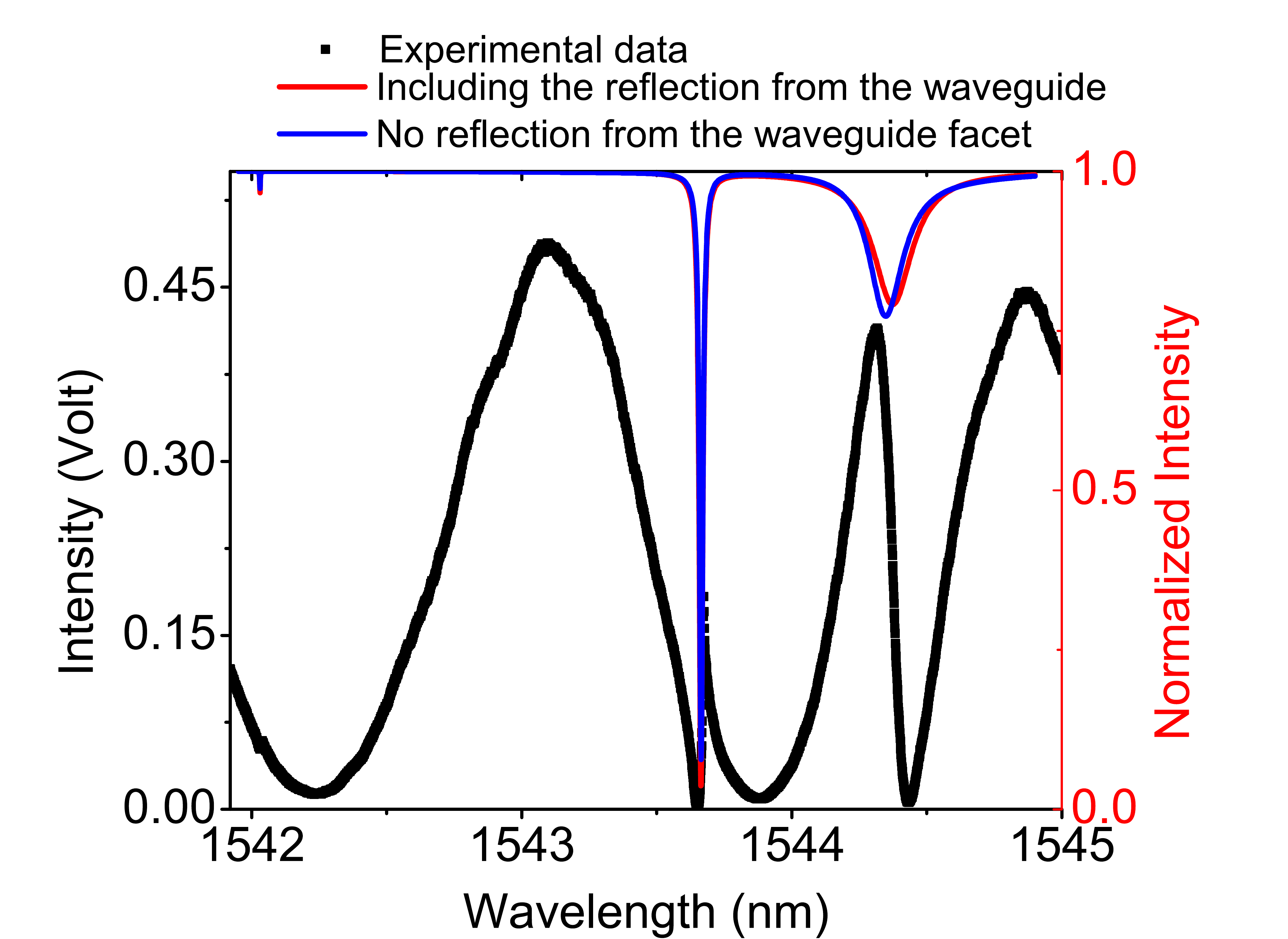}}
\caption{Reflection spectrum of a direct coupled waveguide and multiple-cavity system. Black square: measured spectrum. Red line: calculated spectra taking into account the reflection of the waveguide. Blue line: calculated spectra ignoring reflection of the waveguide facet. The calculations were performed without taking into account the lossy Fabry-Perot cavity between the fiber tip and photonic crystal chip.}
\end{figure}

\section{Theoretical model}

The schematic of the system we consider is shown in Fig. 3. We explicitly include the lensed fiber used to couple light into the waveguide.  Light propagation in this system can be separated into three processes. The first process is the light coupling between the lensed fiber and the input waveguide. The second process is the light transport in the input waveguide.  The last process is the light coupling between the waveguides and cavities. A correct description of the coupling process between the lensed fiber and the input waveguide is essential for the formation the Fabry-P\'erot  fringes in Fig. 2. 

\begin{figure}[H]
\centerline{\includegraphics[width=0.9\columnwidth]{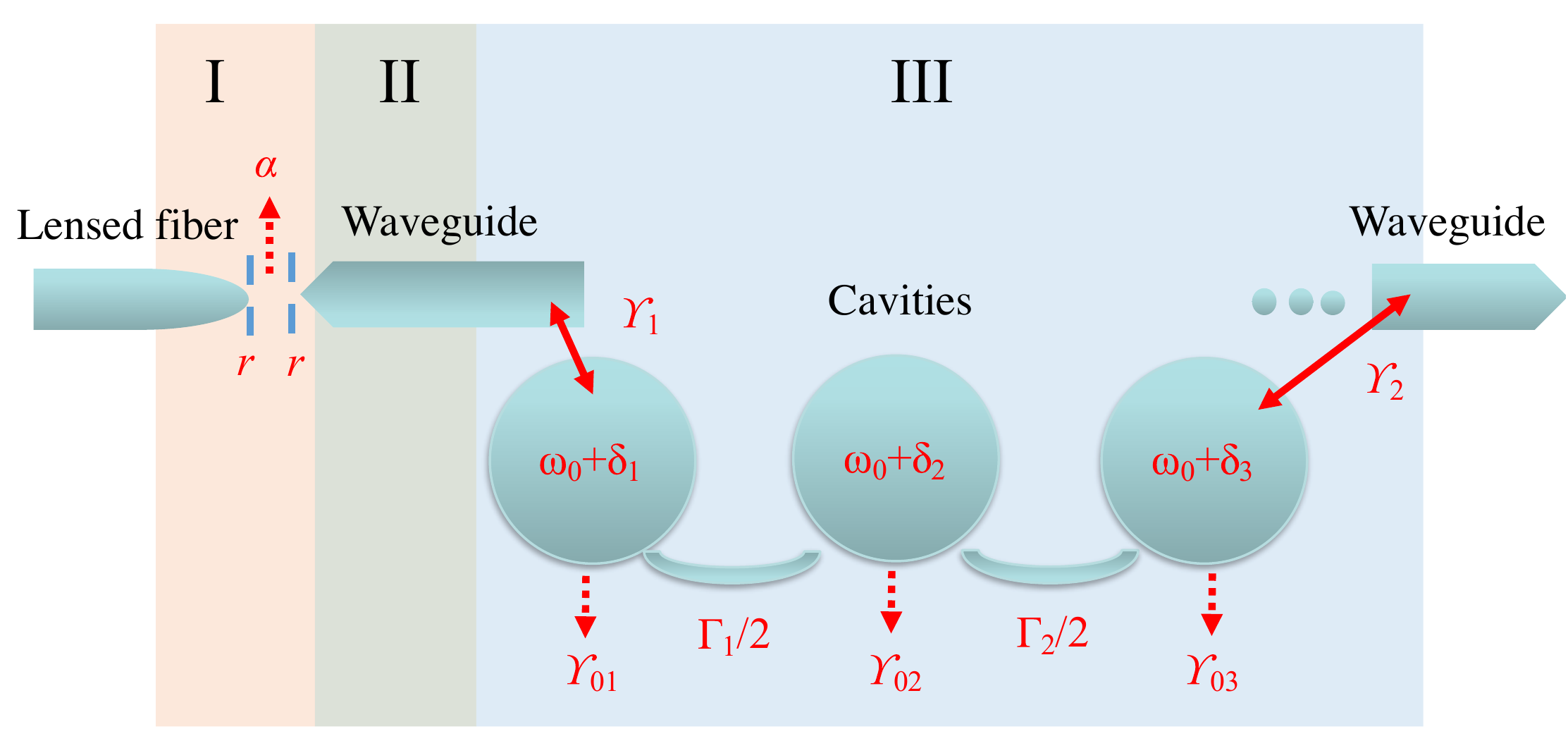}}
\caption{Schematic of the optical system including coupled cavities directly coupled to waveguides. A lensed fiber as a measurement device is also taken into account in the system. }
\end{figure}

We use transfer matrices to model all these processes.  The transfer matrix \cite{Hausbook, Soukoulis} connects the fields of forward and backward propagating waves from the left side to the right side,
\begin{equation}
\begin{pmatrix} S_{\textrm{R}+}  \\ S_{\textrm{R}-}   \end{pmatrix} =  \textbf{M} \begin{pmatrix}  S_{\textrm{L}+}  \\ S_{\textrm{L}-}   \end{pmatrix}.  
\end{equation}
Here $S_{\textrm{R} \pm} $ is the forward (backward) propagating wave on the right side, $S_{\textrm{L} \pm} $ is the forward (backward) propagating wave on the left side and $\textbf{M}$ is the transfer matrix that links them. The matrices describe each process are discussed in the appendix.

\section{Experimental setup and reflection spectra}
In Fig. 1 we show a representation of the sample. Our sample is a photonic crystal membrane structure made of InGaP \cite{alfredoapl2009} with a thickness of 180 nm. The lattice constant is $a=485 \ $nm, the radius of the holes is $0.28a$. There are two waveguides shown in the sample in Fig. 1. One is the input waveguide with the width of 1.1$\sqrt{3}a$. The length of it is 219$a$. The other one is the barrier waveguide, the width of it is $0.98\sqrt{3}a$. In the barrier waveguide, there are three mode-gap cavities \cite{Kuramochi2006}.  They are created by shifting the holes around the barrier waveguide (Fig. 1).  An output waveguide which is in line with the input waveguide is also in the structure. However, it is placed further away from the third cavity.  A polarization maintaining lensed fiber (red cone in Fig. 1) with numerical aperture (NA) 0.55 is used to couple light from a tunable continuous wave (CW) infrared (IR) laser to the sample.  A fiber circulator is used to connect the lensed fiber and laser. The third port of the fiber circulator is connected to a photodiode to measure the reflection spectra of the sample. 

The reflection spectrum is shown in Fig. 4. In Fig. 4(a), we see that the cavity resonances form different lineshapes on top of Fabry-P\'erot fringes. The first Fano resonance (Fig. 4(b)) is between 1544 nm and 1545 nm, and is a wide and deep valley with a slight asymmetry. The second Fano resonance is between 1543.5 nm and 1544 nm, it has a sharp asymmetric line shape. The peak intensity of this resonance is twice smaller than the maximum intensity of the background fringes. The third resonance (Fig. 4(c)) between 1541.95 nm and 1542.10 nm is less pronounced.

\begin{figure}[H]
\centerline{\includegraphics[width=1.02\columnwidth]{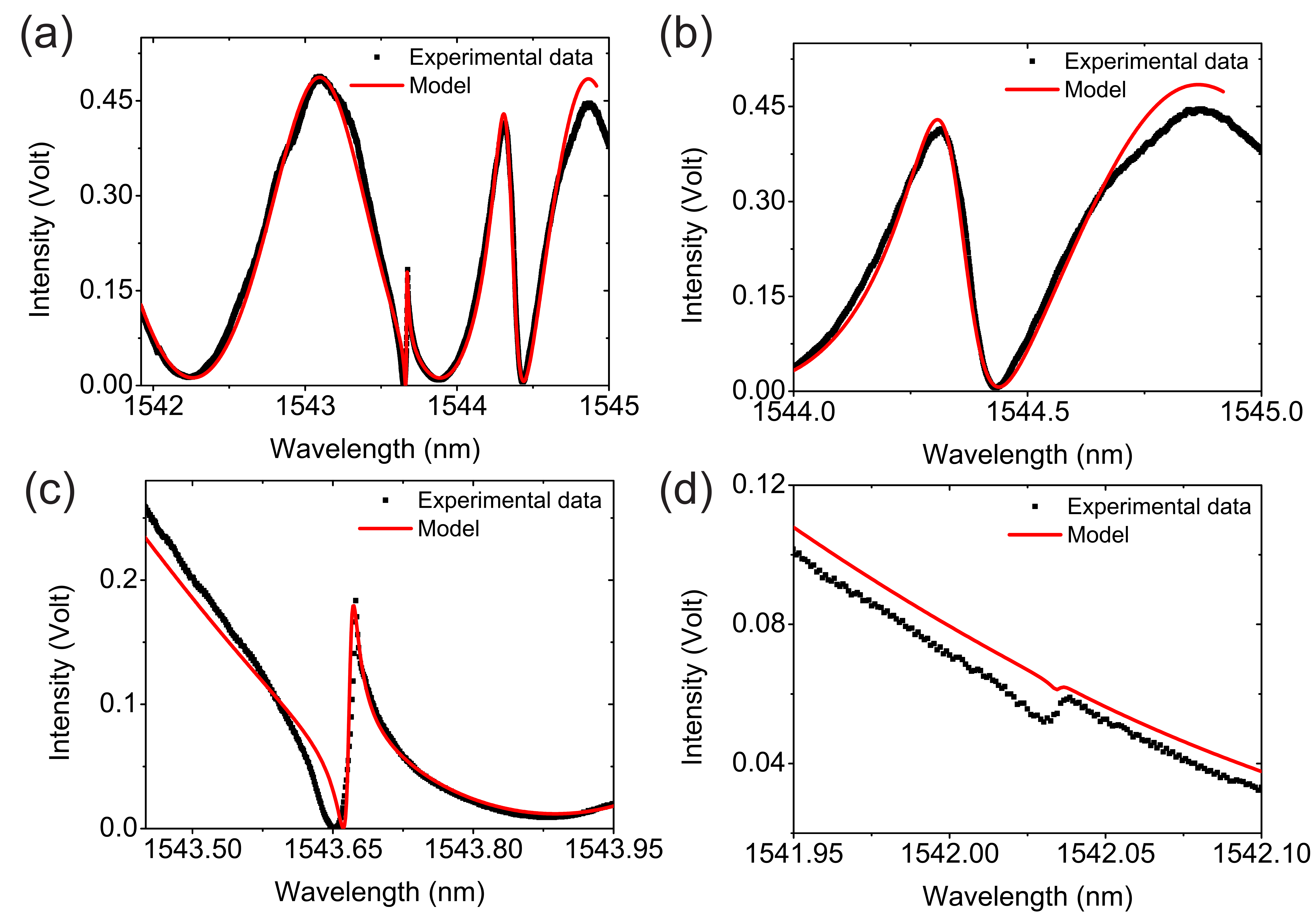}}
\caption{Reference reflection spectrum. The black squares represent the experimental data and the red line is the fit from our theoretical model. (b), (c) and (d) are zoomed in from (a).}
\end{figure}

\begin{figure}[htb]
\centerline{\includegraphics[width=1.02\columnwidth]{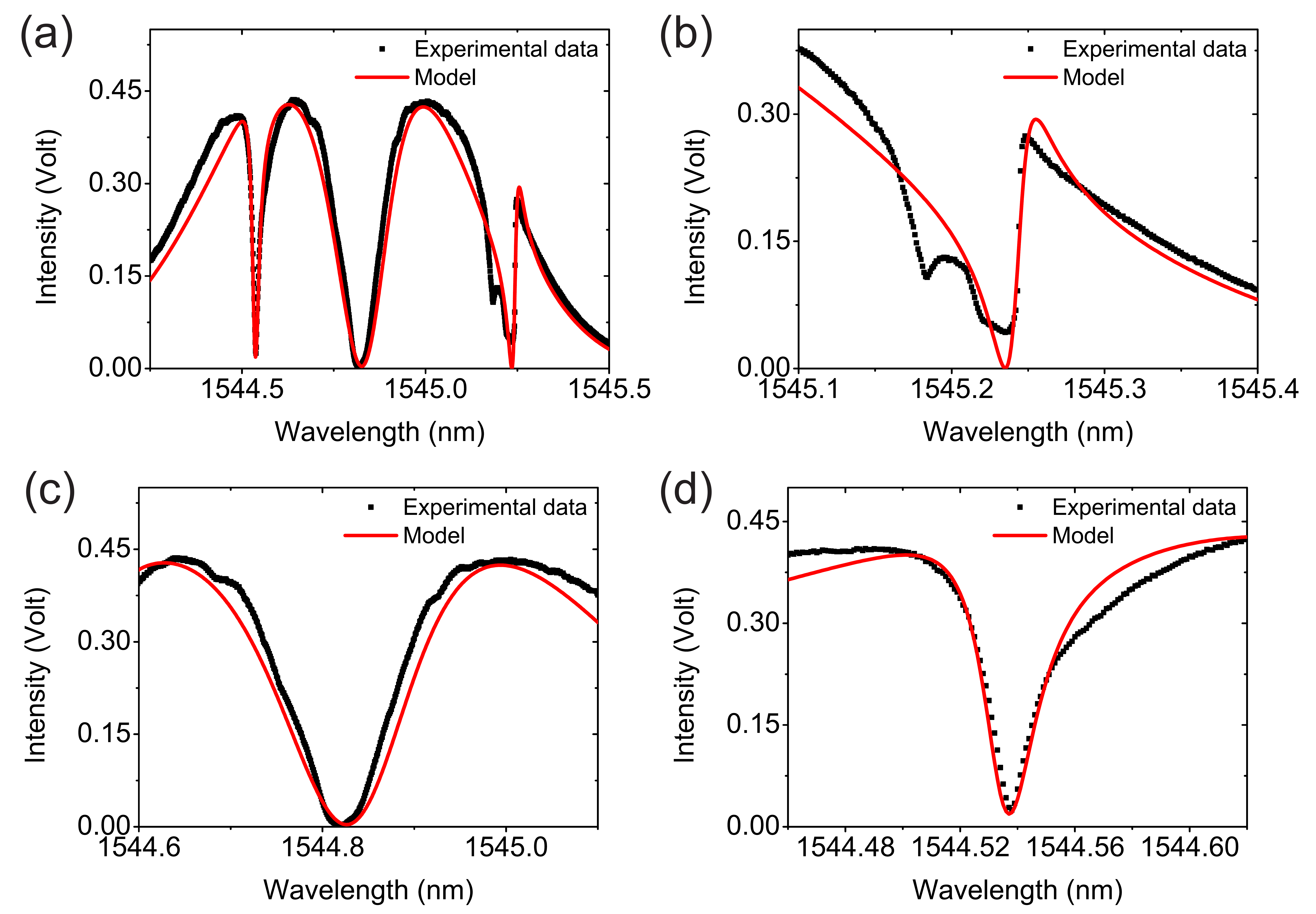}}
\caption{Tuned reflection spectrum. The black squares represent the experimental data and the red line is the fit from our theoretical model. (b), (c) and (d) are zoomed in from (a).}
\end{figure}

In order to cancel the disorder of the cavities \cite{Sokolov2017}, we use a CW diode laser ($\lambda_\textrm{pump}=405 \ $nm) to tune the frequencies of the cavities in the sample by laser induced heating \cite{Sokolov2015}.  To control the first and third cavity simultaneously, two foci are projected on the surface of the sample by an objective with NA 0.4. These foci are generated with the help of a reflective spatial light modulator (SLM) in the pump path. A digital holographic phase pattern is written on the SLM for the generation of the foci \cite{Pasienski08}.  

When the laser spots are focused on the cavities, their resonances are tuned to the red. Although there is no direct laser light on the second cavity, the resonance of cavity 2 also shifts due to heat diffusion.  The reflection spectrum of the tuned device is shown in Fig. 5 \cite{Sokolov2017}. The power of the spots on cavity 1 and cavity 3 is 9 $\mu$W  and 108 $\mu$W respectively. We see that in Fig. 5(a) the three resonances occur within 1 nm in the spectrum. One is with sharp asymmetric line shape. The other two have less asymmetric line shape and form two deep valleys in the spectrum. The modulation of the Fano dip shown in Fig. 6(b) is attributed to SLM noise. 
\section{Analysis of reflection spectra }

The analysis of the reflection spectra is done by fitting experimental results with our model. We first fit the original spectrum, then the spectrum of the tuned device. The values of the parameters in our model are listed in Table 1.

%\begin{table}[ht]
%\begin{center}
%\caption{Types of parameters}
%\begin{tabular}{| c | p{4cm} |  p{4cm} | }
%    \hline
%    Type &  Reference & Pumped \\ \hline
%     Constrained  & $\Gamma_1$, $\Gamma_2$, $\Delta$  & $\Gamma_1$, $\Gamma_2$, $\gamma_1$, $\gamma_{01}$, $\gamma_{02}$,  $\omega_{\textrm{edge}}$\\ \hline
%    Free & $\gamma_1$, $\gamma_{01}$, $\gamma_{02}$, $\gamma_{03}$, $\omega_1$, $\omega_2$, $\omega_3$,  $\omega_{\textrm{edge}}$, $\alpha$, $r$ &  $\gamma_{03}$, $\omega_1$, $\omega_2$, %$\omega_3$, , $\alpha$, $\Delta$\ \\ \hline
%    \end{tabular}   
%\end{center}
%\end{table}

\begin{table}[ht]
\caption{Values of parameters in the theoretical model}
\begin{center}
\begin{tabular}{c  c c }
    \hline \hline
    Parameter                &  Reference             & Tuned                       \\ \hline
     $\Gamma_1/ \omega_0$         &$0.00018 $   &        $0.00018  $  \\ \hline
        $\Gamma_2/ \omega_0$       &$0.00039 $   &        $0.00039  $  \\ \hline
           $\lambda_1 \ $(nm)         &$1544.351 \pm 0.005 $   &        $1544.832 \pm 0.003 $  \\ \hline
              $\lambda_2 \ $(nm)         &$1543.638 \pm 0.005  $   &        $1544.997 \pm 0.003  $  \\ \hline
                $\lambda_3 \ $(nm)         &$1542.08 \pm 0.1 $   &        $1544.764 \pm 0.003 $  \\ \hline
                 $\gamma_{01} / \omega_0$         &$(4.0 \pm 1.0) \times 10^{-6}  $   &   $4.0 \times 10^{-6}  $  \\ \hline
                    $\gamma_{02} / \omega_0$         &$(1.5 \pm 0.8) \times 10^{-6}  $   &       $1.5 \times 10^{-6} $   \\ \hline
                       $\gamma_{03} / \omega_0$         &$ >4 \times 10^{-7} $   &        $(6.80 \pm 3.80) \times 10^{-6} $  \\ \hline
                             $\gamma_1 / \omega_0$         &$ (5.5  \pm 0.2) \times 10^{-5} $   &        $5.5  \times 10^{-5} $  \\ \hline
                             $\omega_{\textrm{edge} } \ (2 \pi c/a)$        &$0.310697 \pm 0.000009  $   &        $ 0.310697  $  \\ \hline  
                                   $m  \ (2 \pi /ac)$        &$0.3149 \pm 0.0005 $   &        $0.3149$    \\ \hline  
                                    $r $        &$0.15  $   &        $0.15$    \\ \hline  
                                     $\alpha$        &$0.795 $   &        $0.795$    \\ \hline  
                                     $\Delta $ (nm)        &$0$   &        $0$    \\ \hline  \hline  
                              \multicolumn{3}{p{8cm}}{$^a$ $\omega_0=a / \lambda_0 \ (2\pi c/a)$ and $\lambda_0$ is 1544.790 nm. $\omega_j= a / \lambda_j \ (j=1, \ 2, \ 3) \ (2\pi c/a)$. The values of $\Gamma_1$ and $\Gamma_2$ have been obtained from our previous work \cite{Sokolov2017}.}  
 \end{tabular}   
 \end{center} 
\end{table}

The fit of our model to the original spectrum is shown in Fig. 4. The fit agrees very well with our data for resonances 1, 2 and the background fringes. It accurately characterizes the period and visibility of the fringes. Meanwhile, it describes the line shapes, widths, heights of the peaks and depths of the trough for resonances 1 and 2. However, we cannot correctly reproduce the depth and width of resonance 3 at the same time. The probable cause for this is a direct coupling term between the waveguide and cavities 2 and 3, or a second neighboring coupling between the cavities. Such terms have been ignored in our model as they would lead to an excessive number of free parameters. 

The fit of our model to the tuned reflection spectrum is shown in Fig. 5. We see that for all three resonances the fit agrees very well with experimental data. There are slight deviations on wings of the resonances between the fit and experimental data. These wings are mostly determined by the Fabry-P\'erot fringes. To characterize them accurately, the accurate knowledge of the curvature of the PhC waveguide band is needed. In our model, we only use two parameters $m$ and $\omega_{\textrm{edge}}$ to describe the band. This is an approximation only valid for a narrow frequency band. The values are obtained by fitting from the reference spectrum shown in Fig 4(a). Thus, the small deviation shown in the tuned spectrum which is in a different wavelength range is reasonable. 

The fact that the fit from our model has an excellent agreement with the experiment shows that our model describes the physical process of the system accurately. It not only explains the physical origin of the observed Fano resonances but also provides the key parameters of the sample such as the intrinsic loss rates of the cavities. The fitting results show that the intrinsic $Q$ factor is larger than $10^5$. After we apply the tuning the frequency detuning between cavity 1 and 2 is reduced below their coupling rate, the same holds for cavities 2 and 3.

\section{Sensitivity of the Fano line shape}
\begin{figure}[H]
\centerline{\includegraphics[width=1.02 \columnwidth]{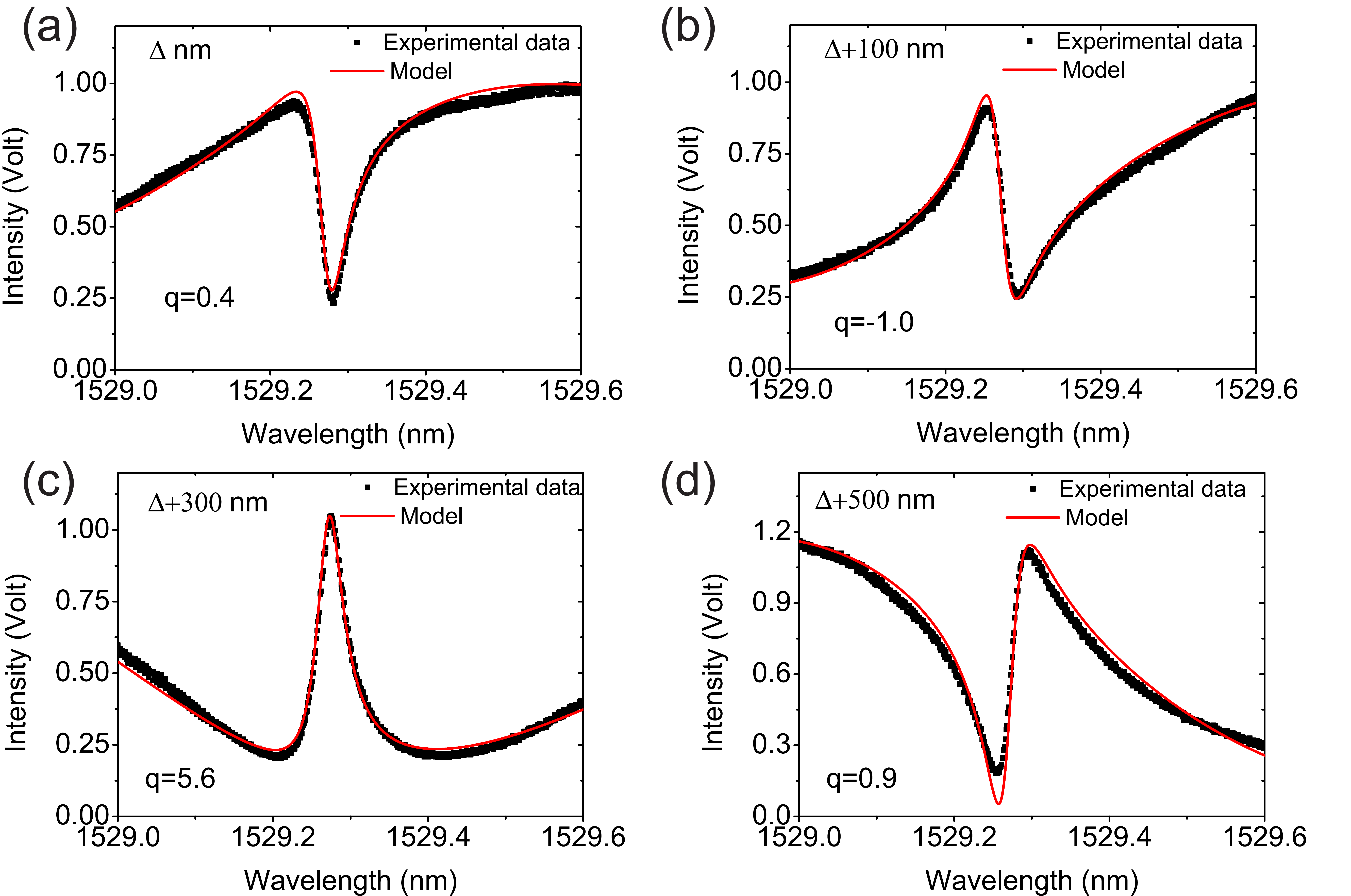}}
\caption{Reflection spectra of a Fano resonance at different air gaps sizes. The black squares represent experimental data and the red lines represent fits from the model. (a) $\Delta$, (b) $\Delta+100 \ $nm, (c) $\Delta+300 \ $nm and (d) $\Delta+500 \ $nm}
\end{figure}

We can describe the lensed fiber, air gap, input waveguide and the cavities together as a special Fabry-P\'erot cavity.  If we ignore the reflection from the facet of the input waveguide, the first ``mirror" of this Fabry-P\'erot cavity is the tip of the lensed fiber, the second ``mirror" of this cavity is the system of photonic crystal cavities.  The length of the cavity is the total length of the air gap together with the input waveguide. The phase shift of a single round trip of this Fabry-P\'erot cavity consists of two parts, the first is the phase shift from propagation and the second is the phase shift due to the reflection from the second ``mirror".  The specialty of the second ``mirror" is that it is very dispersive around the cavity frequencies.  The Fano line shape is determined by the phase shift of the round trip. Therefore, we conclude that the Fano lineshape of the resonance can be tuned by changing the length of the air gap.

We perform an experiment to test this prediction on a new sample with the same parameters as our previous sample. Due to the inevitable disorder, the resonance positions of the cavities appear at different wavelengths. We measure reflection spectra with different sizes of the air gap. This was done by moving the sample step by step away from the lensed fiber with our precise translation stage. 

The measured reflection spectra are shown in Fig. 6.  In Fig. 6(a), the reference spectrum is presented (the reference distance between the lensed fiber and sample is denoted by $\Delta$ and is the distance where the coupling is optimized), and we see a Fano resonance is in the form of a dip with a bit asymmetry.  After the reference measurement, we increase the length of the air gap with a step size of 100 nm. The spectra with air gap sizes $\Delta+100 \  $nm, $\Delta+300 \ $nm and $\Delta+500 \ $nm are shown in Fig. 6(b), Fig. 6(c) and Fig. 6(d) respectively. In Fig. 6(b), we see a sharp asymmetric Fano resonance with a peak at short wavelength and a dip at a longer wavelength. In complete contrast to the reference spectrum, we see a peak with a slight asymmetry instead of a dip in Fig. 6(c). In Fig. 6(d), we again see a sharp asymmetric Fano resonance, however, it is almost a flipped version of Fig. 6(b). It has a dip at a short wavelength and peak at a longer wavelength. We also retrieve the q parameter that describes the asymmetry of Fano lines using the Fano line formula \cite{Miroshnichenko2010}.  We also plot the fits from our analytical model. In the fits, all further parameters are kept the same for the results from Fig. 6(a) to Fig. 6(d) except the coupling loss, since it increases slightly as we increase the size of the air gap. % For the size of the air gap, we gradually change its value according to what shown in Fig. 6. 
The fits agree well with our experiment, which confirms that changing the air gap size causes the drastic change of the Fano line shape. The results shown in Fig. 6 confirm our prediction that the shape of a Fano resonance can be tuned by changing the size of the air gap between the sample and lensed fiber, and demonstrate that the origin of the Fano line shape is indeed the interference of the sharp resonances with the broad resonances defined by the reflection of the lensed fiber and coupling loss.  The maximum distance we move the fiber to manipulate the Fano line shape is only 500 nm from the optimal coupling point. Since the Rayleigh range is  1.9 $\mu$m, the coupling efficiency only experiences a very small change.  On the contrary, the Fano line shape as we show in Fig. 6 experiences a drastic change.  

\section{Conclusion}
In summary, Fano resonances in the reflection spectra of a direct-coupled waveguide-cavities system in a photonic crystal membrane structure have been experimentally and theoretically investigated. Our theoretical model has an excellent agreement with our experimental results and provides important information on the very low bare loss rate of the cavities. The origin of the Fano lineshape is the interference between the wave reflected from the lensed fiber and the wave reflected from the photonic crystal cavities. The path length difference between these waves is a round trip of the air gap size and the input waveguide. We propose and experimentally show that the Fano asymmetric parameter can be tuned drastically by only changing the air gap size between the sample and the fiber by 100 nm which is a number well below the Rayleigh range.  Our model can be used to investigate other physical processes in the system, such as the dynamical tuning of the Fano asymmetry by ultra-fast switching \cite{Yuce13,Nozaki2010}. 

% ---------------------------------------------------------------- 

\section*{Acknowledgments} \label{sec:Acknowledgement}
The authors thank Sanli Faez, Pritam Pai and Willem L. Vos for helpful discussions and Cornelis Harteveld for technical support. This work is supported by the European Research Council project No. 279248. APM acknowledges a Vici grant from the Nederlandse Organisatie voor Wetenschappelijk Onderzoek.

\appendix
\renewcommand{\theequation}{\Alph{section}.\arabic{equation}}
\numberwithin{equation}{section}
\
\section{Transfer Matrices of the theoretical model}
\subsection{Process I: coupling between the measurement device and the waveguide}
In the process of coupling between the measurement device (lensed fiber) and the waveguide, there are a few physical events we have to consider.  The first is the reflection at the boundaries between the lensed fiber and air, and between air and the facet of the input waveguide. The second is the coupling loss. The third is the light propagation in air. 

Reflection at the boundaries can be modeled as reflection from partially reflecting mirrors. The transfer matrix of a partially reflecting mirror has the form as \cite{Hausbook},
\begin{equation}
\textbf{M}_\textrm{pr} = \frac{1}{i\sqrt{1-r^2}} \begin{pmatrix} -1 & -r  \\ r & 1 \end{pmatrix}, 
\end{equation}
where $r$ is the amplitude reflection coefficient. In the model here for simplicity, we assume it is the same for both boundaries. 

The propagation of light in the air gap is modeled as,
\begin{equation}
\textbf{M}_\textrm{Air} =  \begin{pmatrix} \exp {(i \frac{\omega}{c} \Delta) } & 0  \\ 0 & \exp {(-i \frac{\omega}{c} \Delta)} \end{pmatrix}, 
\end{equation}
where $\Delta$ is the length of the air gap.

The corresponding physical process of the coupling loss can be modeled as,
\begin{equation}
\textbf{M}_\textrm{loss} =  \begin{pmatrix} \sqrt{1-{\alpha}} & 0  \\ 0 & \frac{1}{\sqrt{1-{\alpha}} } \end{pmatrix},
\end{equation}
where $\alpha$ is the power loss in the coupling process.

The matrix that describes process I is the multiplication of the matrices discussed above, it is 
\begin{equation}
\textbf{M}_{\textbf{I}}= \textbf{M}_\textrm{pr} \cdot \textbf{M}_\textrm{loss} \cdot \textbf{M}_\textrm{Air} \cdot \textbf{M}_\textrm{pr}
\end{equation}

\subsection{Process II: propagation in the photonic crystal waveguide}
The propagation in the photonic crystal waveguide has the same form as Eq. (3). However, the dispersion of the waveguide has to be considered.  The matrix is 
\begin{equation}
\textbf{M}_{\textbf{II}} =  \begin{pmatrix} \exp {ik_b(\omega)L_l} & 0  \\ 0 & \exp {-ik_b(\omega)L_l} \end{pmatrix}  \quad (l=1,2),
\end{equation}
where the dispersive quantity $k_b(\omega)$ is the Bloch wave vector of the wave in the waveguide and  $L_l$ is the length of waveguide $l$.  It is known that close to the band edge,  the dispersion of the photonic crystal can be expressed as $\omega=\omega_{\textrm{edge}}+(k_b-\pi/a)^2/(2m)$ \cite{Baron2015, Lian15}, where $\omega$ is the frequency of the waveguide mode, $ \omega_{\textrm{edge}}$ is the frequency of the band edge of the waveguide mode,  and $m=(\partial^2\omega / \partial^2 k)^{-1}$ is the effective photon mass close to the band edge. As a result, we can express the Bloch wave vector $k_b$ as $k_b=\pi/a-\sqrt{2m(\omega-\omega_{\textrm{edge}})}$. 

\subsection{Process III:  light coupling between the waveguides and cavities.}

To derive the matrix which describes the light coupling between the waveguides and cavities, we use the temporal coupled equations. In the equations,  we only consider the coupling between the first (last) cavity and input (output) waveguide, with coupling rates $\gamma_1$ and $\gamma_2$ respectively.  We use $a_j(t) \ (j=1, 2, 3) $ to denote the time evolution of the field in cavity $j$, and  $S_{l\pm}$ $(l = 1, 2)$ to denote the amplitude of the mode in waveguide $l$. $l=1$ represents the input waveguide and $l=2$ represents the output waveguide. The "$\pm$" represents forward (backward) propagation. From coupled mode theory \cite{Haus1991,mitbook, Khan1999, Vuckovic2005}, we use following equations to describe the dynamics of the system,
\begin{equation}
\left\{ \begin{aligned}
     &  \frac{da_1}{dt}=i\omega_1a_1-(\gamma_{01}+\gamma_1) a_1+ \sqrt[]{2\gamma_1}S_{1+}+i\frac{\Gamma_1}{2}a_2  \\
       & \frac{da_2}{dt}=i\omega_2a_2-\gamma_{02} a_1+i\frac{\Gamma_1}{2}a_1+i\frac{\Gamma_2}{2}a_3 \\
         & \frac{da_3}{dt}=i\omega_3a_3-(\gamma_{03}+\gamma_2) a_1+\sqrt[]{2\gamma_2}S_{2-}+i\frac{\Gamma_2}{2}a_2   \\
                  &  S_{1-}=-S_{1+}+\sqrt[]{2\gamma_1}a_1 \\
    &  S_{2-}=-S_{2+}+\sqrt[]{2\gamma_2}a_3
\end{aligned}
\right.
\end{equation}

In Eq. (A.6), $\omega_j$ is the actual bare frequency of cavity $j$ which is defined as $\omega_j=\omega_0+\delta \omega_j$ where $\omega_0$ is the intended intrinsic frequency of the cavities, and $\delta \omega_j \ (j=1,2,3)$ representing the  frequency deviation of cavity $j$ from  the actual bare resonance frequency due to fabrication disorder. $\Gamma_1$ is the  coupling rate between cavity 1 and 2, $\Gamma_2$ is the coupling rate between cavity 2 and 3.
We solve Eq. (A.6) in a matrix equation form in the Fourier domain, and obtain matrix $\textbf{M}_{\textbf{III}}$ which is the transfer matrix that links  $(S_{2-}, \ S_{2+})$ and $(S_{1-}, \ S_{1+})$.  The lengthy but straightforward expression is not shown here.
 
The matrix that describes all the processes is $\textbf{M}_\textrm{sys}=\textbf{M}_{\textbf{III}} \cdot \textbf{M}_{\textbf{II}} \cdot \textbf{M}_{\textbf{I}}$.  In our experiment the value of $\gamma_2$ is small enough to assume that all the elements after the output waveguide decouple from the system and do not influence the reflection spectrum.  

\subsection{Model without the lensed fiber}
In the model without the lensed fiber, $\textbf{M}_{\textbf{I}}=\textbf{M}_\textrm{pr}$. The rest process described by  $\textbf{M}_{\textbf{II}}$ and $\textbf{M}_{\textbf{III}}$  remain the same.

\end{document}